\documentclass[a4paper,11pt,reqno]{article}

\usepackage{a4wide}
\usepackage{amsmath,amsfonts,amssymb}
\usepackage[english]{babel}
\usepackage{soul}
\usepackage{nicefrac}
\usepackage[mathscr]{euscript}
\usepackage{setspace}
\usepackage{datetime}
\usepackage[sort&compress,merge,numbers]{natbib}

\usepackage{mathrsfs}
\usepackage[T1]{fontenc}
\usepackage{mathpazo}

\usepackage[breaklinks=true]{hyperref}

\numberwithin{equation}{section}

\hypersetup{
			colorlinks=true,
			urlcolor=blue,
			citecolor=magenta,
			linkcolor=blue,
     }

\let\oldsqrt\sqrt
\def\sqrt{\mathpalette\DHLhksqrt}
\def\DHLhksqrt#1#2{%
\setbox0=\hbox{$#1\oldsqrt{#2\,}$}\dimen0=\ht0
\advance\dimen0-0.2\ht0
\setbox2=\hbox{\vrule height\ht0 depth -\dimen0}%
{\box0\lower0.4pt\box2}}

\author{
  \begin{minipage}{.97\linewidth}
    \vspace{.95cm}
       \begin{center}
      \begin{small}
                \textbf{Anastasios C. Petkou},$^{1,2}$
     \textbf{P. Marios Petropoulos}$^{3,4}$ and 
      \textbf{Konstantinos Siampos}$^5$
              \end{small}
    \end{center}
    \vspace{0.3cm}
    \hspace{1.3cm}
\begin{minipage}{.7\linewidth}
\begin{center}     {\it \begin{footnotesize}
\hbox{\kern-1.cm\vbox{\vskip0cm
 \begin{itemize}
               \item[$^1$] Department of Physics\\ 
  Institute of Theoretical Physics\\
  Aristotle University of Thessaloniki\\ 
  54124 Thessaloniki, Greece
                 \vskip0.3cm
             \item[$^3$] Centre de Physique Th\'eorique\\ 
        Ecole Polytechnique, CNRS UMR 7644\\ 
        Universit\'e Paris-Saclay\\
        91128 Palaiseau Cedex, France
               \vskip0.33cm
      \end{itemize}}
\kern-5cm\vbox{\vskip0cm
\begin{itemize}
               \item[$^2$] Dipartimento di Fisica\\
   Universit\`a di Roma ``Tor Vergata'' \\
Via della Ricerca Scientifica\\
00133 Rome, Italy
 \vskip0.3cm
 \item[$^4$] Laboratoire de Physique Th\'eorique et Hautes Energies\\ 
        Sorbonne Universit\'es, CNRS UMR 7589\\ 
        UPMC Paris 6\\
        4 place Jussieu, 75005 Paris, France 
       \end{itemize}
       \vskip0.05cm
}}
     \end{footnotesize}}
\end{center}
    \end{minipage}
    \vspace{0.3cm}\begin{minipage}{.7\linewidth}
\begin{center}     
{\it \begin{footnotesize}
\hbox{\kern3.7cm\vbox{\vskip0cm
 \begin{itemize}
               \item[$^5$] Albert Einstein Center for Fundamental Physics\\
Institute for Theoretical Physics\\ 
Bern University\\
Sidlerstrasse 5, 3012 Bern, Switzerland
      \end{itemize}}
}
     \end{footnotesize}}\end{center}
    \end{minipage}
    \vspace{0.3cm}
     \begin{small}
  \begin{center}
 \texttt{petkou@physics.auth.gr, marios@polytechnique.edu, siampos@itp.unibe.ch}
 \end{center}
 \end{small}
  \end{minipage}
}

\title{\vspace{3.cm}
 \boldmath  \begin{Large}
    \textbf{Geroch group for Einstein spaces and holographic integrability}
 \end{Large}
   \unboldmath
}

\begin{document}

\begin{titlepage}
\maketitle
\thispagestyle{empty}

 \vspace{-15.cm}
  \begin{flushright}
  CPHT-RR046.0914
    \end{flushright}
 \vspace{13cm}

\begin{center}
\textsc{Abstract}\\  
\vspace{1cm}	
\begin{minipage}{1.0\linewidth}

We review how Geroch's reduction method is extended from Ricci-flat to Einstein spacetimes. The Ehlers--Geroch $SL(2,\mathbb{R})$ group is still present in the three-dimensional sigma-model that captures the dynamics, but only a subgroup of it is solution-generating. Holography provides an alternative three-dimensional perspective to integrability properties of Einstein's equations in asymptotically anti-de Sitter spacetimes. These properties emerge as conditions on the boundary data (metric and energy--momentum tensor) ensuring that the hydrodynamic derivative expansion be resummed into an exact four-dimensional Einstein geometry. The conditions at hand are invariant under a 
set of transformations dubbed holographic $U$-duality group. The latter fills the gap left by the Ehlers--Geroch group in Einstein spaces, and allows for solution-generating maps mixing e.g. the mass and the nut charge.

\end{minipage}
\end{center}


\end{titlepage}

\onehalfspace

\vspace{-1cm}
\begingroup
\hypersetup{linkcolor=black}
\tableofcontents
\endgroup
\noindent\rule{\textwidth}{0.6pt}


\section*{Introduction}
\addcontentsline{toc}{section}{Introduction}

Einstein's equations are generically not integrable. Under some assumptions, the system possesses  integrability properties often revealed  by solution-generating techniques. 
In 1970, Geroch exhibited in Refs. \cite{Geroch, Geroch:1972yt} a method for generating vacuum solutions of Einstein's equations, generalizing previous work by Ehlers \cite{Ehlers}. 
The starting point was a four-dimensional manifold $\mathcal{M}$, endowed with a metric $\mathrm{g}$ and vanishing Ricci tensor. A generic Killing vector $\xi$ was also assumed for $\mathrm{g}$, with scalar twist $\omega$ and norm $\lambda$. A coset space $\mathcal{S}$ was further constructed  as the quotient of $\mathcal{M}$ by  the one-parameter group generated by
$\xi$.
The core of the proposed method was to set a one-to-one mapping between  $\mathcal{S}$ and  $\mathcal{M}$, and recast the four-dimensional Einstein's equations in terms of the data on  $\mathcal{S}$: the metric $\mathrm{h}$  on the projected space $\mathcal{S}$
orthogonal to $\xi$,
and the scalar fields $\lambda$ and $\omega$. Any new triplet $( \mathrm{h}',  \omega',\lambda')$ satisfying that set of equations could be promoted to another, potentially new four-dimensional  vacuum solution $\mathrm{g}'$ with one isometry.

Geroch's sigma-model is not integrable but exhibits a continuous group of symmetries, the $U$-duality group, and allows to map any solution into another. Although this mapping is non-local amongst the uplifted four-dimensional solutions, it is local at the level of the three-dimensional data. Indeed, keeping the metric $\mathrm{h}$ within the conformal class of fixed $\lambda \mathrm{h}$, new solutions can be generated as M\"obius transformations  of $\tau=\omega +i \lambda$: $ \tau'=\nicefrac{a\tau +b}{c\tau+d}$ with 
$\left(\begin{smallmatrix}      a & b \\ c & d  \end{smallmatrix}\right)$ in $SL(2,\mathbb{R})$. 
In concrete examples such as Schwarzschild--Taub--NUT solutions with mass $m$ and nut charge $n$, the compact  
subgroup  of elements $\left(\begin{smallmatrix}      \cos \nicefrac{\psi}{2} & \sin \nicefrac{\psi}{2} \\ -\sin \nicefrac{\psi}{2} & \cos \nicefrac{\psi}{2}  \end{smallmatrix}\right) \in SO(2)\subset SL(2,\mathbb{R})$ induces rotations of angle $\psi$ in the parameter space $(m,n)$: $m+in \to \text{e}^{-i\psi}(m+in)$. 
Non-compact transformations $\left(\begin{smallmatrix}      a & b \\ 0 & \nicefrac{1}{a}  \end{smallmatrix}\right)\in N \subset SL(2,\mathbb{R})$ act homothetically:
$(m,n)\to (\nicefrac{m}{a} , \nicefrac{n}{a} )$.

The above techniques have been extensively developed over the years for vacuum or electrovacuum solutions (see e.g. \cite{Alekseev:1980ew,Mazur:1983,Alekseev:2010mx}). In particular, following Ernst \cite{Ernst:1967wx, Ernst:1968}, Geroch's method generalizes when 2 commuting Killing vectors are available on $(\mathcal{M}, \text{g})$ and allows to reduce Einstein's dynamics to a two-dimensional sigma-model. The latter possesses full affine symmetry\footnote{This is the algebra customary quoted as \emph{Geroch} in the literature, whereas it is common to name the finite-dimensional algebra  \emph{Ehlers}.}  and is integrable \cite{Breitenlohner:1986um, belinskii, Maison:1978es, maison2, Mazur:1982} (see also \cite{Bernard:2001pp, Bardoux:2013swa} for more references). Further generalizations of the method have been studied in great detail within supergravity theory, in various dimensions \cite{Alekseev:2004zz,Alekseev:2008gh,Figueras:2009mc,Mishima:2005id,Iguchi:2007is}. Such analyses provide an important and  complementary perspective with respect to Geroch's algebraic solution-generating technique.

Geroch's approach is more intricate in the presence of a cosmological constant, and only a few sparse examples
were available till recently in the framework of Ernst's equations \cite{Charmousis:2006fx,Caldarelli:2008pz,Astorino:2012zm}. Actually, even though Geroch insisted in starting with a vacuum solution $(\mathcal{M}, \mathrm{g})$, all the requirements necessary to translate Einstein's equations to three-dimensional terms remain valid in the more general case of Einstein spaces: $\lambda $ and $\omega$ are well-defined and together with the coset $(\mathcal{S}, \mathrm{h})$, they provide a complete characterization of $(\mathcal{M}, \mathrm{g})$. This has led the authors of Ref. \cite{Leigh:2014dja} to 
reconsider the original analysis. Besides the three-dimensional fields $ (\omega,\lambda)$, it is necessary to introduce an extra field $\kappa$ playing the r\^ole of the conformal factor, and study the dynamics of 
$(\kappa, \omega, \lambda)$ in the sigma-model. Although the action of the $SL(2,\mathbb{R})$ is again well defined, the $U$-duality group turns out to be only a subgroup of it, and the benefit for generating new solutions is limited. In particular, this group does not include the generator which realizes the mapping  $m+in \to \text{e}^{-i\psi}(m+in)$. This is unfortunate because a full  Schwarzschild Taub--NUT solution does exist on AdS, and it is legitimate to ask whether this is indeed a manifestation of a hidden symmetry, as for the vacuum case, that would allow to obtain the nut charge as an image of the mass parameter.

The advent of gauge/gravity holographic correspondence allows to recast the above questions regarding 
Einstein spaces in a slightly different manner. Originally formulated at the microscopic level, holography has been extended macroscopically as a relationship between gravity plus matter in asymptotically (locally) anti-de Sitter spaces and some phenomenological boundary conformal field theory in one dimension less. The latter is usually a macroscopic quantum state, which might -- but needs not to -- be 
in the hydrodynamic regime. This is how fluids emerge in holography.

A bulk Einstein space  $(\mathcal{M}, \text{g})$, delivers a boundary space $\mathcal{B}$ with boundary metric $\text{g}_{\text{bry.}}$ and boundary energy--momentum tensor $\text{T}$.  
The two holographic pieces of data for pure-gravity bulk dynamics are thus
$\text{g}_{\text{bry.}}$  and $\text{T}$. These data allow in turn to reconstruct the bulk Einstein space using  e.g. the Fefferman--Graham series  expansion \cite{PMP-FG1, PMP-FG2}. For arbitrary boundary data, this series expansion is generically non-resummable and the bulk solution is not exact. 

It is possible to set up integrability requirements, which if obeyed by the boundary data, ensure the resummability of the expansion at hand, and an exact Einstein space at the end \cite{Gath:2015nxa, Petropoulos:2015fba, Mukhopadhyay:2013gja}. 
The proposed set of integrability requirements can be viewed as a kind of \emph{integrable holographic reduction} of Einstein's equations, alternative to Geroch's sigma model. It turns out that this set of integrability requirements leaves some freedom to map $(\text{g}_{\text{bry.}}, \text{T})$ onto $(\text{g}_{\text{bry.}}', \text{T}')$. Hence, the reconstructed  $(\mathcal{M}, \text{g}')$ from $(\text{g}_{\text{bry.}}', \text{T}')$  leads to a possibly new exact Einstein space.  Schematically, a holography-based solution-generating  technique works as follows:
\begin{equation}
\nonumber
(\mathcal{M}, \text{g})  \underset{r\to \infty}{\longrightarrow}  (\mathcal{B}, \text{g}_{\text{bry.}}, \text{T}) \underset{U_\text{hol.}}{\rightarrow}(\mathcal{B}, \text{g}_{\text{bry.}}', \text{T}') \underset{\text{exact reconstruction}}{\longrightarrow}  (\mathcal{M}, \text{g}'),
\end{equation}
where $r$ is the holographic radial coordinate and $U_\text{hol.}$ the \emph{holographic $U$-duality group}.
Both $(\text{g}_{\text{bry.}}, \text{T})$ and $(\text{g}_{\text{bry.}}', \text{T}')$ are boundary data satisfying the set of integrability requirements. In the present scheme, holography provides an alternative solution-generating technique for Einstein spaces.

What makes it possible for revealing integrability conditions for boundary data and further discovering the seed for a holographic $U$-duality group, is the hydrodynamic interpretation of the boundary state. 
The purpose of this note is to report on recent progress \cite{ Caldarelli:2012cm, Mukhopadhyay:2013gja, Petropoulos:2014yaa,Gath:2015nxa, Petropoulos:2015fba} (see also Ref. \cite{Berkeley:2015mmc})  made in using holographic fluids for understanding integrable corners of Einstein's equations. Prior to that and in the spirit of a comprehensive presentation, we also report on the generalization of Geroch's method in Einstein spaces, as was exposed in \cite{Leigh:2014dja}. This will actually be our starting point, reviewed in Sec. \ref{geroch}.
We will then move in Sec. \ref{der-res} to the integrability properties in the framework of the derivative  expansion, which is an alternative to the Fefferman--Graham expansion, inspired by the  black-brane paradigm and proposed in \cite{Haack:2008cp, Bhattacharyya:2008jc, Bhattacharyya:2008ji}. The emergence of the holographic $U$-duality transformations will also be discussed, in general as well as for a particular class of integrable boundary data, where  the r\^ole of the transformation as a rotation in the mass--nut plane is transparent. This shows that \textit{the holographic $U$-duality group fills indeed the gap left by the Geroch group in Einstein spaces}.

\section[Following Geroch]{Following Geroch\footnote{We reproduce in this section results borrowed from \cite{Leigh:2014dja}.}}\label{geroch}

\subsection{From four to three dimensions}

We assume $\mathcal{M}$ a four-dimensional manifold endowed with a Lorentzian-signature metric $\mathrm{g}=g_{ab}\text{d}x^a \text{d}x^b $, and $\xi$ a Killing field. The latter has norm 
$\lambda=\left\| \xi\right\|^2$
and twist one-form\footnote{Here ``$\star$'' is the four-dimensional Hodge duality performed with $\eta_{abcd}=\sqrt{-\det g}\, \epsilon_{abcd}$ ($\epsilon_{0123}=1$).
We also recall that for $\zeta$ a vector,  
$\text{i}_\zeta$ is the contraction with $\zeta$. 
\label{magicform}}  
$w=-2 \text{i}_\xi \star \text{d}\xi$.
The Killing one-form $\xi$ obeys identically
\begin{equation}\label{Kilprop}
\star \text{d}\star\text{d}\xi = 2  \text{i}_\xi \mathrm{Ric},
\end{equation}
where $\mathrm{Ric}$ is the four-dimensional Ricci tensor. Assuming the metric be Einstein
\begin{equation}\label{Eins-g}
\mathrm{Ric}=\Lambda \mathrm{g},
\end{equation}
Eq. \eqref{Kilprop} reads:
\begin{equation}\label{Kilprop-Eins}
\text{d}\star \text{d}\xi = 2  \Lambda \star \xi.
\end{equation}
It is straightforward to show that 
$\text{d}w= 0$, which allows to define the scalar twist locally as
\begin{equation}
w=\text{d}\omega.
\end{equation}

We can define the space $\mathcal{S}$ as a quotient of $\mathcal{M}$
with respect to the action of the one-parameter group generated by $\xi$. As $\xi$ may not be hypersurface-orthogonal (which would imply zero twist), this coset space needs not be a subspace of $\mathcal{M}$. There is a natural metric $\text{h}$ on $\mathcal{S}$ induced by $\text{g}$ of $\mathcal{M}$ as
\begin{equation}\label{met-h}
h_{ab}=g_{ab}-\frac{\xi_a\xi_b}{\lambda},
\end{equation}
which defines the projector onto $\mathcal{S}$ is 
\begin{equation}\label{proj-h}
h^b_a=\delta^b_a-\frac{\xi^b\xi_a}{\lambda}.
\end{equation}
For the metric \eqref{met-h}, the volume form and the fully antisymmetric tensor read:
\begin{equation}\label{vol-h}
\text{Vol}_{\text{h}} =  \frac{1}{\sqrt{-\lambda}}\text{i}_\xi \text{Vol}_{\text{g}}
\Leftrightarrow
 \eta_{abc}= \frac{- 1}{\sqrt{-\lambda}}\eta_{abcd}\xi^d,
\end{equation}
where we assumed for concreteness $\lambda<0$ so that the Killing vector $\xi$ is timelike and $\text{h}$ is spatial (the whole reduction procedure goes smoothly through when $\lambda$ is positive).

Following \cite{Geroch}, let us quote some basic features of the geometrical relationship between $\mathcal{M}$ and $\mathcal{S}$.
There is a natural one-to-one correspondence between tensors on $\mathcal{S}$ and tensors $\text{T}$ on $\mathcal{M}$, transverse and invariant with respect to the Killing flow
\emph{i.e.} that satisfy $\text{i}_\xi \text{T}=0$ and $\mathcal{L}_\xi \text{T}=0$.
Assume now a tensor $\text{T}$ on $\mathcal{S}$. It is easy to show that the covariant derivative $\mathscr{D}$ defined following this correspondence, 
\begin{equation}\label{covd-h}
\mathscr{D}_c T_{a_1\ldots a_p}^{\hphantom{a_1\ldots a_p}b_1\ldots b_q} =
h_c^\ell h_{a_1}^{m_1}\dots  h_{a_p}^{m_p}h^{b_1}_{n_1}\dots  h^{b_q}_{n_q}
\nabla_\ell T_{b_1\ldots b_p}^{\hphantom{b_1\ldots b_p}n_1\ldots n_q} 
\end{equation}
with $\nabla$ the Levi--Civita derivative on $\mathcal{M}$, coincides with the unique Levi--Civita covariant derivative on $\mathcal{S}$. This provides a Riemann tensor on $\mathcal{S}$ in terms of the Riemann tensor of $\mathcal{M}$:
\begin{equation}\label{R-h}
\mathscr{R}_{abcd}
= h_{[a}^{\hphantom{[}p}h_{b]}^{q}
h_{[c}^{\hphantom{[}r}h_{d]}^{s}
\left(R_{pqrs}+\tfrac{2}{\lambda}\left(
\nabla_p\xi_q
\nabla_r\xi_s
+\nabla_p\xi_r
\nabla_q\xi_s
\right)\right)
\end{equation}
(the calligraphic  letters refer to $\mathcal{S}$ tensors). These equations  are more general than Gauss--Codazzi equations, since $\xi$ needs not be hypersurface orthogonal.

The existence of a Killing vector $\xi$ on $(\mathcal{M},\mathrm{g})$ allows us to recast the dynamics of $\mathrm{g}$ in terms of $(\mathrm{h},\omega,\lambda)$, which can all be regarded as fields on $\mathcal{S}$. For that, 
one extracts the  $\mathcal{S}$-Ricci tensor $\mathscr{R}_{ab}$ from \eqref{R-h} and further determines the  $\mathcal{S}$-Laplacians of $\lambda$ and $\omega$. The final equations are, under the assumption \eqref{Eins-g}, 
\begin{equation}\label{Ger-eq}
\begin{array}{rcl}
\mathscr{R}_{ab}&=&
\frac{1}{2\lambda^2}
\left(
\mathscr{D}_a \omega\mathscr{D}_b \omega-h_{ab}
\mathscr{D}^c \omega\mathscr{D}_c \omega
\right)
+\frac{1}{2\lambda}\mathscr{D}_a\mathscr{D}_b\lambda
 -\frac{1}{4\lambda^2}\mathscr{D}_a\lambda\mathscr{D}_b\lambda
 + \Lambda h_{ab},  
\\
\mathscr{D}^2\lambda&=&\frac{1}{2\lambda}\left(
\mathscr{D}^c \lambda\mathscr{D}_c \lambda-2\mathscr{D}^c \omega\mathscr{D}_c \omega
\right)-2 \lambda\Lambda,\\
\mathscr{D}^2\omega&=&\frac{3}{2\lambda}
\mathscr{D}^c \lambda\mathscr{D}_c \omega.
\end{array}
\end{equation}

Equations \eqref{Ger-eq} provide in principle new solutions $(\mathrm{h}',\omega',\lambda')$, which 
can be promoted to a new metric $\text{g}'$ with symmetry $\xi'$ following the procedure described in Sec. \ref{3to4}. Most importantly, these equations can be recast in a useful manner by introducing a three-dimensional reference metric $\hat{\text{h}}$, defined as 
\begin{equation}\label{newmet3}
h_{ab}=\frac{\kappa}{\lambda}\hat h_{ab}.
\end{equation}
The dilaton-like field $\kappa$ captures one of the degrees of freedom carried by the metric $\text{h}$, and inheritates its dynamics from the latter's.
This is useful for probing mini-superspace solutions with frozen $\hat{\text{h}}$, because it allows for one gravity degree of freedom to remain dynamical, together with  $\omega$ and $\lambda$. It should be emphasized here that the original Geroch's reference metric was defined as $\tilde h_{ab}={\lambda}h_{ab}$. Freezing $\tilde{\text{h}}$ removes all $\text{h}$ degrees of freedom, leaving only  $\omega$ and $\lambda$ as dynamical fields.  In the presence of a cosmological constant, the scalar degree of freedom $\kappa$  is crucial for the system to capture e.g. mass and nut parameters simultaneously. This is explicitly shown when performing the mini-superspace analysis of the integrability properties of 
\eqref{Ger-eq}, assuming that the space $\mathcal{S}$ 
is topologically $\mathbb{R}\times \mathcal{S}_2$, and the metric $\hat{\text{h}}$ of the form
$
\mathrm{d}\hat{s}^2=\mathrm{d}\sigma^2+\mathrm{d}\Omega^2,
$
where $\mathrm{d}\Omega^2$ is the two-dimensional $\sigma$-independent piece. 
Details can be found in the original publication \cite{Leigh:2014dja}.
 
Introducing $\tau=\omega + i \lambda$ together with \eqref{newmet3},
Eqs. \eqref{Ger-eq} read:
\begin{eqnarray}
\hat{\mathscr{R}}_{ab}&=&
-\tfrac{2}{(\tau-\bar\tau)^2}    
\hat{\mathscr{D}}_{(a}\tau\, \hat{\mathscr{D}}_{b)}\bar\tau
+\tfrac{1}{2\kappa}\left(
\hat{\mathscr{D}}_a \hat{\mathscr{D}}_b \kappa+\hat  h_{ab}
\hat{\mathscr{D}}^c \hat{\mathscr{D}}_c \kappa
\right)\nonumber\\
&&-\tfrac{1}{4\kappa^2}\left(3 \hat{\mathscr{D}}_{a}\kappa \hat{\mathscr{D}}_{b}\kappa
 +\hat  h_{ab} \hat{\mathscr{D}}^c\kappa \hat{\mathscr{D}}_c\kappa\right)
 +4i \Lambda\tfrac{\kappa}{\tau-\bar\tau}\hat h_{ab},  
\label{eq-hab}\\
\hat{\mathscr{D}}^2\tau&=&\tfrac{2}{\tau-\bar \tau}
\hat{\mathscr{D}}^c \tau\hat{\mathscr{D}}_c \tau-\tfrac{1}{2\kappa}\hat{\mathscr{D}}^c \kappa\hat{\mathscr{D}}_c \tau
-2i \Lambda\kappa,\label{eq-tau}
\end{eqnarray}
where all hatted quantities refer to the metric $\hat{\text{h}}$.
These equations describe the dynamics of the fields $(\hat{\text{h}},\kappa, \tau)$, the equation for $\kappa$ being the trace of \eqref{eq-hab}:
\begin{equation}\label{eq-kap}
\hat{\mathscr{D}}^2\kappa=
\tfrac{3}{4\kappa} \hat{\mathscr{D}}^{c}\kappa \hat{\mathscr{D}}_{c}\kappa
+
\tfrac{\kappa}{(\tau-\bar\tau)^2}    
\hat{\mathscr{D}}^{c}\tau\, \hat{\mathscr{D}}_{c}\bar\tau
-6i \Lambda\tfrac{\kappa^2}{\tau-\bar\tau}
+\tfrac{\kappa}{2}
\hat{\mathscr{R}} .
\end{equation}

Equations \eqref{eq-hab}, \eqref{eq-tau} and \eqref{eq-kap} can be obtained by extremizing, with respect to $\hat h_{ab}, \tau$ and $\kappa$, the action  
$
S = \int_{\mathcal{S}} \mathrm{d}^3 x \sqrt{\hat h}\mathcal{L}
$
with Lagrangian density
\begin{equation}\label{gen-L-den}
\boxed{
\mathcal{L}=-\sqrt{-\kappa}\left(\frac{\hat{\mathscr{D}}^a\kappa
\hat {\mathscr{D}}_a\kappa}{2\kappa^2}+2
\frac{\hat {\mathscr{D}}^a\tau
\hat {\mathscr{D}}_a\bar \tau}{(\tau-\bar \tau)^2}+
\hat{\mathscr{R}}-4i \Lambda\frac{\kappa}{\tau-\bar\tau}
\right).}
\end{equation}
As usual, not all components of Eqs. \eqref{eq-hab} are evolution equations: there are also the Hamiltonian 
and momentum constraints. The sigma-model action at hand describes a system of ``matter'' fields $\kappa, \omega$ and  $\lambda$, 
which define a three-dimensional target space with metric  
\begin{equation}
\label{st-geo-n}
\mathrm{d}s^2_{\text{target}}
=\sqrt{-\kappa}\left(-
\frac{\mathrm{d}\kappa^2}
{\kappa^2}
+\frac{\mathrm{d}\omega^2+\mathrm{d}\lambda^2}{\lambda^2}\right),
\end{equation}
interacting through a ``matter'' potential, 
\begin{equation}
\mathcal{V}=\sqrt{-\kappa}\left(\hat{\mathscr{R}}-2 \Lambda\frac{\kappa}{\lambda}
\right).\label{st-geo-pot}
\end{equation}
It is customary to introduce in this potential the three-dimensional gravity contribution $\sqrt{-\kappa}\hat{\mathscr{R}}$, as ultimately, in a mini-superspace analysis, the field  $\hat h$ is frozen.

\subsection{From three to four dimensions}\label{3to4}

We would like to remind here how any new solution $(\mathrm{h}',\omega',\lambda')$ of Geroch's Eqs. \eqref{Ger-eq}, 
can be uplifted without ambiguity to a new metric $\text{g}'$ with symmetry $\xi'$ on  the four-dimensional manifold $\mathcal{M}$. The procedure is based on the following observation: the two-form defined on $\mathcal{S}$ as\footnote{Here $\star^3_{\text{h}'}$ stands for the three-dimensional Hodge-dual with respect to $\text{h}'$.}
\begin{equation}
\text{F}'= \frac{1}{(-\lambda')^{\nicefrac{3}{2}}}
\star^3_{\text{h}'}\text{d}\omega'
\end{equation}
is \emph{closed} and thus, locally exact
\begin{equation}\label{eta}
\text{F}'=\text{d}\eta'.
\end{equation}
The field $\eta'$, \emph{a priori } defined on $\mathcal{S}$, can be promoted to a field on $\mathcal{M}$ by adding the necessary exact piece such that its normalization is 
\begin{equation}\label{eta-norm}
\text{i}_\xi \eta'=1,
\end{equation}
and this defines a new Killing field on $\mathcal{M}$
\begin{equation}\label{newkil}
\xi'= \eta' \lambda'. 
\end{equation}
The new four-dimensional metric finally reads:
\begin{equation}\label{newmet}
g_{ab}'=h_{ab}'+\frac{\xi_a'\xi_b'}{\lambda'}.
\end{equation}

\subsection{Integrability}\label{integr}
The target-space metric \eqref{st-geo-n} is conformal to $\mathbb{R}\times H_2$, which has an  $\mathbb{R}\times SL(2,\mathbb{R})$ isometry group generated by 
\begin{equation}\label{U1k}
\zeta=\frac{1}{2}\kappa\partial_\kappa,
\end{equation}
and
\begin{equation}\label{Lsl}\
\xi_+=\partial_\omega, \quad
\xi_-=\left(\lambda^2-\omega^2\right)\partial_\omega-2\omega\lambda\partial_\lambda, \quad
\xi_2=\omega\partial_\omega+\lambda\partial_\lambda,
\end{equation}
obeying 
\begin{equation}
\left[\xi_+,\xi_- \right]=-2\xi_2,\quad
\left[\xi_+,\xi_2 \right]=\xi_+,\quad
\left[\xi_2,\xi_- \right]=\xi_-.\label{Lslt}
\end{equation}
For the metric \eqref{st-geo-n}, $\zeta$ is a conformal Killing field, whereas the $\xi$s remain Killing, and generate the $SL(2,\mathbb{R})$ Geroch (sometimes called Ehlers) group. The quadratic Casimir of the latter is generated by the Killing tensor $\Xi= -\nicefrac{\kappa}{\lambda^2}\left(\mathrm{d}\omega^2+\mathrm{d}\lambda^2\right)$.

For non-vanishing $\Lambda$, the potential  $\mathcal{V}$ breaks part of the $SL(2,\mathbb{R})$ symmetry, and only $\xi_+$ leaves it invariant. This allows for a single conservation law. Actually, another one turns out to be available for $\xi_2$ thanks to the Hamiltonian constraint (see details in \cite{Leigh:2014dja}).

The system described by \eqref{gen-L-den} is not generically integrable. The Ehlers--Geroch $SL(2,\mathbb{R})$ group is not large enough to account for the infinite number of conserved charges, necessary for integrability. It allows nevertheless to map any solution onto another one, and is thus solution-generating. The situation is more interesting when the original Einstein space possesses two commuting Killing vectors. In this case, the three-dimensional sigma model can be further reduced to two dimensions \cite{Ernst:1967wx, Ernst:1968}, with potentially remarkable integrability properties.

We can summarize the various situations as follows:
\paragraph{\boldmath $\Lambda=0$ \unboldmath} Any solution $(\hat{\text{h}},\kappa, \tau)$ provides another solution $(\hat{\text{h}},\kappa, \tau')$ with 
\begin{equation}
\tau'=\frac{a\tau +b}{c\tau+d}\, ,\quad
\begin{pmatrix}      a & b \\ c & d  \end{pmatrix} \in SL(2,\mathbb{R}). 
\end{equation}
In the presence of an extra commuting Killing vector, the corresponding two-dimensional sigma model is invariant under a full $\hat{sl}(2,\mathbb{R})$ affine algebra. The latter guarantees integrability \cite{Breitenlohner:1986um,belinskii, Maison:1978es, maison2, Mazur:1982}. Actually, the two-Killing vacuum solutions are known and classified according to their Petrov type: the Pleba\'nski--Demia\'nski family is type-D and the Kundt solutions are type-N or type-III (see e.g. \cite{PMP-GP}).

\paragraph{\boldmath $\Lambda\neq 0$ \unboldmath} In this case, the situation is more intricate. Firstly, only a subgroup $N \subset SL(2,\mathbb{R})$ is solution-generating \cite{Leigh:2014dja}: from the Einstein space $(\hat{\text{h}},\kappa, \tau)$  with $\Lambda$, we obtain another Einstein space $(\hat{\text{h}},\kappa, \tau')$ with\footnote{The potential \eqref{st-geo-pot} is indeed invariant under the $\omega$ $b$-shift generated by $\xi_+$  (see \eqref{Lsl}); it is also invariant under $\xi_2$ (parameter $a$) provided $\Lambda$ gets transformed along with $\lambda=\mathrm{Im}\, \tau$.}
\begin{equation}
\label{solgen-GL}
\Lambda' = a^2 \Lambda\quad \text{and}
\quad
\tau'=a\left(a\tau +b\right) ,\quad
\begin{pmatrix}      a & b \\ 0 & \nicefrac{1}{a}  \end{pmatrix} \in N\subset SL(2,\mathbb{R}). 
\end{equation}
Secondly, assuming an extra commuting  Killing vector, the obtained two-dimensional sigma model has never been proven to be integrable or to posses an affine symmetry. We nevertheless expected it to be, as it is admitted that all two-Killing Einstein spaces with cosmological constant are known (again Pleba\'nski--Demia\'nski and Kundt spaces). 

In conclusion, although for vanishing $\Lambda$ the Geroch approach provides the appropriate framework for discussing solution-generating techniques and integrability, it seems that it fails to achieve that for $\Lambda \neq 0$, despite the common available formalism and the close relationship amongst known Ricci-flat and Einstein solutions. This has led to reconsider the latter case from a holographic perspective.

\section[Holographic integrability]{Holographic integrability}
 \label{der-res}

\subsection{The derivative expansion and the resummed Einstein space} \label{der-exp}

For pure gravity dynamics, holography is equivalent to a Hamiltonian evolution. It requires two pieces of fundamental holographic data:  the boundary metric and the boundary energy--momentum tensor. Any set of such data, allow in principle to reconstruct an Einstein bulk geometry, captured either in the Fefferman--Graham or in the derivative expansions. 

In holographic terms, integrability of Einstein's equations translates into the resummability of the series expansion, which reconstructs the bulk from boundary data. In this perspective, the natural question in the search of integrable sectors for Einstein's equations is therefore as follows: given a class of boundary metrics, what are the conditions it should satisfy, and which energy--momentum tensor should it be accompanied with in order for an \emph{exact} dual bulk Einstein space to exist? A more difficult companion question is whether the integrability conditions also ensure regularity.

From the perspective of physics, and putting aside the questions regarding regularity, these issues have been thoroughly investigated in Refs. \cite{Caldarelli:2012cm, Mukhopadhyay:2013gja, Petropoulos:2014yaa, Gath:2015nxa,Petropoulos:2015fba}, and we will here review the results. From a mathematical viewpoint the related \emph{filling-in problem} was studied long ago, and has been a guide in our approach. A round three-sphere is a positive-curvature, maximally symmetric Einstein space with $SU(2)\times SU(2)$ isometry. A hyperbolic four-space $H_4$ is a negative-curvature, maximally symmetric Einstein space, which is a foliation over round three-spheres. In this sense, the round three-sphere  is filled-in with $H_4$, the latter being the only regular metric filling-in this three-dimensional space. 

The natural question to ask in view of the above is how to fill-in the more general Berger sphere $S^3$,  which is  a homogeneous but non-isotropic deformation of the round sphere. LeBrun studied the filling-in problem
in general terms \cite{PMP-LeBrun82}, and showed that an analytic three-metric can be regularly filled-in by a four-dimensional Einstein space that has self-dual (or anti-self-dual) Weyl tensor, \emph{i.e.} by a \emph{quaternionic} space. In modern holographic words, LeBrun's result states that requiring regularity makes the boundary metric a sufficient piece of data for reconstructing the bulk. Regularity translates into conformal self-duality, which effectively reduces by half the independent Cauchy data of the problem.

From LeBrun's filling-in problem we learn that integrability conditions involve the structure of the Weyl tensor, and are closely related to regularity conditions. In spacetimes with Lorentzian signature, Weyl self-duality cannot be the answer, as this would trivialize the geometry, leading to conformally flat spacetimes. 
As we will see, however, the Weyl tensor remains the central object in integrability, perceived as the resummability of an expansion. Whether our integrability conditions also guarantee regularity, remains an open question.

In the subsequent analysis, we will be using the derivative expansion. The latter assumes the existence of a null   
geodesic congruence in the bulk, defining tubes that extend from the boundary inwards. 
On the boundary, this congruence translates into a timelike congruence, and the aforementioned derivative series expansion is built on increasing derivative order of this field. 

Weyl covariance is the guideline for the reconstruction of spacetime  based on the derivative expansion \cite{Bhattacharyya:2008jc, Bhattacharyya:2008ji, Haack:2008cp}: the bulk geometry should be insensitive to a conformal rescaling of the boundary metric $\text{d}s^2_{\text{bry.}}\to  \nicefrac{\text{d}s^2_{\text{bry.}}}{{\cal B}^2}$. 
Covariantization with respect to rescalings requires to introduce a Weyl connection one-form:
\begin{equation}
\label{Wcon}
\text{A}=\text{a} -\frac{\Theta}{2} \text{u}  ,
\end{equation}
where $ \text{u} $ is tangent to the timelike normalized ($u_\mu u^\mu=-1$) congruence and $\text{a}$ is its acceleration. The Weyl connection 
transforms as $\text{A}\to\text{A}-\text{d}\ln {\cal B}$. Ordinary covariant derivatives $\nabla$ are thus traded for Weyl-covariant ones $\mathscr{D}=\nabla+\theta\,\text{A}$ with $\theta$ the conformal weight of the tensor under consideration.

In the present analysis, we will be interested in situations where the boundary congruence $\text{u}$ is shearless. 
Vanishing shear simplifies considerably the reconstruction of the asymptotically locally AdS bulk geometry because it reduces the available Weyl-invariant terms. As a consequence, at each order of $\mathscr{D}\text{u}$, the terms compatible with  Weyl covariance of the bulk metric are nicely organized. Even though we cannot write them all at arbitrary order, the structure of the first orders suggest that resummation, whenever possible, should lead to the following \cite{Bhattacharyya:2008jc, Bhattacharyya:2008ji, Caldarelli:2012cm, Mukhopadhyay:2013gja,
Gath:2015nxa, Petropoulos:2015fba,Petropoulos:2014yaa}:
\begin{equation}
\text{d}s^2_{\text{res.}} =
-2\text{u}(\text{d}r+r \text{A})+r^2k^2\text{d}s^2_{\text{bry.}}+\frac{\Sigma}{k^2}
+ \frac{\text{u}^2}{\rho^2} \left(\frac{3 T_{\mu\nu}u^\mu u^\nu}{k \kappa }r+\frac{C_{\mu \nu}u^\mu \eta^{\nu\rho\sigma}\omega_{\rho\sigma}}{2k^6}\right).
\label{papaefgenres}
\end{equation}
Here $r$ the radial coordinate whose dependence is explicit, 
$x^\mu$ are the three boundary coordinates extended to the bulk, on which depend implicitly the various functions,
$\kappa=\nicefrac{3k}{8\pi G}$ and $k$ is a constant related to the bulk cosmological constant as $\Lambda=-3k^2$. We have furthermore introduced the following objects: 
\begin{itemize}
\item the conserved boundary energy--momentum tensor  $\text{T}=T_{\mu\nu}\text{d}x^\mu\text{d}x^\nu$, appearing via the boundary energy density
\begin{equation}
\label{Piprop}
\varepsilon(x)=T_{\mu\nu}u^\mu u^\nu;
\end{equation}
\item the vorticity $\omega$ of the boundary timelike congruence ($\omega =\nicefrac{1}{2}\left(\mathrm{d}\mathrm{u} +
\mathrm{u} \wedge\mathrm{a}\right)=\nicefrac{1}{2}\,\omega_{\mu\nu}\, \mathrm{d}\mathrm{x}^\mu\wedge\mathrm{d}\mathrm{x}^\nu$) together with its dual vector of components  $q u^\mu = \eta^{\mu\nu\sigma}\omega_{\nu\sigma}$;
\item the boundary metric $\text{d}s^2_{\text{bry.}}=g_{\mu\nu}\text{d}x^\mu \text{d}x^\nu$ and its conserved Cotton tensor\footnote{Reminder: the components of the Cotton tensor are $C^{\mu\nu}=\eta^{\mu\rho\sigma}
\nabla_\rho \left(R^{\nu}_{\hphantom{\nu}\sigma}-\frac{R}{4}\delta^{\nu}_{\hphantom{\nu}\sigma} \right),
$ with $\eta_{\mu\rho\sigma}$ given by a formula similar to the four-dimensional one of footnote \ref{magicform} with $\epsilon_{012}=1$.} $\text{C}=C_{\mu\nu}\text{d}x^\mu\text{d}x^\nu$
for which it is customary to introduce
\begin{equation}
\label{cofx}
c(x) = C_{\mu\nu}u^\mu u^\nu;
\end{equation}
\item the boundary tensor 
\begin{equation}
\label{sigma}
\Sigma=
\Sigma_{\mu\nu} 
\text{d}x^\mu\text{d}x^\nu=-2\text{u}\mathscr{D}_\nu \omega^\nu_{\hphantom{\nu}\mu}\text{d}x^\mu- \omega_\mu^{\hphantom{\mu}\lambda} \omega^{\vphantom{\lambda}}_{\lambda\nu}\text{d}x^\mu\text{d}x^\nu
-\text{u}^2\frac{\mathscr{R}}{2}
\end{equation}
with
\begin{eqnarray}
\mathscr{D}_\nu\omega^{\nu}_{\hphantom{\nu}\mu}&=&\nabla_\nu\omega^{\nu}_{\hphantom{\nu}\mu},
\\
\mathscr{R}&=&R +4\nabla_\mu A^\mu- 2 A_\mu A^\mu;  \label{curlR}
\end{eqnarray}
\item the function $\rho$, defined as
\begin{equation}\label{rho2}
 \rho^2=r^2 +\frac{1}{2k^4} \omega_{\mu\nu} \omega^{\mu\nu} = r^2 +\frac{q^2}{4k^4},
\end{equation}
which performs the resummation as the derivative expansion is manifestly organized in powers of $q^2=2 \omega_{\mu\nu} \omega^{\mu\nu}$.
\end{itemize}
Under a conformal rescaling of the boundary metric, the tensor $\Sigma$ is invariant, while $\text{C}\to {\cal B}\, \text{C}$, and at the same time $\text{T}\to {\cal B}\, \text{T}$, $\text{u}\to \nicefrac{\text{u}}{{\cal B}}$ (velocity one-form) and $\omega\to \nicefrac{\omega}{{\cal B}}$. 

The important issue  regarding the resummed bulk metric \eqref{papaefgenres} is to set up the boundary conditions 
under which this metric is an exact Einstein space. This has been discussed in detail in the already quoted literature and we will report our findings in Sec. \ref{rescon}. Meanwhile, we should pause and mention an important result: \emph{whenever the resummed metric $\text{d}s^2_{\text{res.}}$ is Einstein, it is  Petrov algebraically special}. The proof of this statement\footnote{The interested reader can find all the technicalities of that proof in App. A of Ref. \cite{Petropoulos:2015fba}.} is based on the shear-free nature of the congruence $\text{u}$. The absence of shear for this boundary timelike congruence guarantees that the corresponding null  bulk congruence $\partial_r$, which is geodesic, is also shear-free. Thanks to the Goldberg--Sachs theorem and its generalizations, a reconstructed Einstein bulk geometry \eqref{papaefgenres}
is algebraically special \emph{i.e.} of Petrov type II, III, D, N or O. Furthermore, the bulk congruence $\partial_r$ provides a \emph{principal null direction}.

We would like to insist on the r\^ole played by the absence of shear for the boundary congruence. Not only this assumption allows to discard the large number of Weyl-covariant tensors available when the shear is non-vanishing, which would have probably spoiled any resummation attempt; but it also selects the algebraically special geometries, known to be related with integrability properties. One can safely say that the absence of shear is intimately related with the resummability of the derivative expansion. 
Although we cannot exclude that some exact Einstein type I space might be successfully reconstructed from boundary data, or that no exact resummation involves a congruence with shear, this looks very unlikely in view of the above. 

\subsection{The resummability conditions} \label{rescon}

We will list in this section all boundary ingredients and conditions needed for reaching holographically exact bulk Einstein spacetimes when using the derivative expansion, organized around the derivatives of the boundary congruence $\text{u}$. This expansion assumes small derivatives, small curvature, and small higher-derivative curvature tensors for the boundary metric. This limitation is irrelevant for us since we are ultimately interested in resumming the series, allowing even for non-perturbative (\emph{i.e.} non-hydrodynamic) contributions.

At the perturbative level, the fluid interpretation is applicable and the boundary timelike congruence is identified with the boundary fluid velocity field. This is how the holographic fluid emerges. 
Beyond the perturbative framework, however, the fluid interpretation is not faithful due to the presence of non-hydrodynamic modes in the boundary energy--momentum tensor. In that case, the boundary timelike congruence refers only to the hydrodynamic part of the energy--momentum tensor and this raises an aside, although important question in our analysis: given the boundary data $\text{d}s^2_{\text{bry.}}$ and 
$\text{T}$, what is the congruence $\text{u}$ that will organize the expansion? This congruence cannot be read off from the energy--momentum tensor, since the non-hydrodynamic pieces of the latter blur the fluid interpretation. It turns out that this question has a very simple answer, which comes as part of the resummability ansatz and requirements. These are organized as follows:
\begin{enumerate}
\item \underline{The boundary metric.} 

We consider a three-dimensional boundary spacetime, equipped with a metric $\text{d}s^2_{\text{bry.}}=g_{\mu\nu}\text{d}x^\mu \text{d}x^\nu$ ($\mu, \nu, \ldots = 0,1,2$). 
Given a generic three-dimensional metric, there is a unique way to express it as a fibration over a conformally flat two-dimensional base:\footnote{See e.g. \cite{Coll} and App. A of Ref. \cite{Petropoulos:2015fba}.
In the presence of isometries, expression \eqref{PDbdymet} may not be unique. 
Notice that we could set $\Omega=1$, without spoiling the generality -- as we are interested in the conformal class.} 
\begin{equation}
\label{PDbdymet}
\text{d}s^2_{\text{bry.}}=-\Omega^2(\text{d}t-\text{b})^2+\frac{2}{k^2P^2}\text{d}\zeta\text{d}\bar\zeta,
\end{equation}
with $P$ and $\Omega$ arbitrary real functions of $(t,\zeta, \bar \zeta)$, 
and
\begin{equation}
\label{frame}
\text{b}=B(t,\zeta, \bar \zeta)\, \text{d}\zeta+\bar B(t,\zeta, \bar \zeta)\, \text{d}\bar\zeta.
\end{equation}
\item \underline{The boundary timelike shear-free congruence.}  

In three-dimensional geometries, 
there is basically a unique timelike, normalized and shearless congruence.\footnote{This statement is not true in the presence of isometries, where more shearless congruences may exist. In these cases, the distinct congruences are equivalent.}  When the metric is of the form \eqref{PDbdymet}, this congruence is precisely the comoving one:
\begin{equation}
\label{ut}
\text{u}= -\Omega(\text{d}t-\text{b}).
\end{equation}
This defines our fluid congruence.
 
\item \underline{The boundary reference tensors.}

As quoted in Sec. \ref{der-exp}, the resummed bulk metric \eqref{papaefgenres}, when Einstein, is Petrov algebraically special. The Petrov classification is obtained from the eigenvalue equation for the Weyl tensor. 
In particular, the Weyl tensor and its dual can be combined in a pair of complex-conjugate (self-dual and anti-self-dual) tensors.
Each of these tensors has two pairs of bivector indices, which can be used to deliver a complex two-index tensor. Its components are naturally packaged inside a complex symmetric $3 \times 3$ matrix $\text{Q}$ with zero trace (see e.g. \cite{Stephani:624239} for this construction).
This matrix encompasses the ten independent real components of the Weyl tensor and the associated eigenvalue equation determines the Petrov type.

For a general Einstein space, the leading-order ($\nicefrac{1}{r^3}$) coefficient $\text{S}^{\pm}$ in the Fefferman--Graham expansion of the complex Weyl tensor $\text{Q}^{\pm}$ exhibits a specific combination of the components of the boundary Cotton and energy--momentum tensors \cite{Mansi:2008br, Mansi:2008bs}.
Indeed, performing a similarity transformation with $\text{P}={\rm diag}(\pm i,-1,1)$, we obtain 
$\text{T}^\pm = -\text{P} \, \text{S}^{\pm}\text{P}^{-1}$ with components
\begin{equation}
	\label{eqn:Tref}
	T_{\mu\nu}^\pm = T_{\mu\nu} \pm \frac{i\kappa }{3k^3 }C_{\mu\nu}.
\end{equation}
We refer to $\text{T}^\pm $ as the \emph{reference energy--momentum tensors} because they play the r\^ole of a complex-conjugate pair of fictitious conserved boundary sources, always accompanying the boundary geometry. 
These tensors must be of a canonical form because the bulk resummed metric is algebraically special. A canonical form is dictated by the Segre classification (see e.g. \cite{Chow:2009km}): it can be either \emph{perfect-fluid, pure-radiation or pure-matter}. 

\item \underline{The resummability conditions.}

The reference tensors given in Eq.~\eqref{eqn:Tref} are by construction symmetric, traceless and conserved:
\begin{equation}
\label{Tref-cons}
\boxed{
\nabla\cdot \text{T}^\pm=0.}
    \end{equation}
Choosing a specific form for these  tensors, and assuming a boundary metric $\text{d}s^2_{\text{bry.}}$, we are led to two conditions. The first, provides a set of equations that the boundary metric must satisfy:
\begin{equation}
\label{C-con}
\text{C}=\frac{3 k^3}{\kappa}  \text{Im} \text{T}^+.
    \end{equation}
The second delivers the boundary energy--momentum tensor it should be accompanied with for an exact bulk ascendent spacetime to exist:
\begin{equation}
\label{T-con}
\text{T}= \text{Re} \text{T}^+.
    \end{equation}
 \end{enumerate}

The above presentation of the boundary ingredients and requirements can thus be summarized as follows:  a generic boundary metric accompanied with canonical reference tensors $\text{T}^\pm$ (perfect-fluid, pure-matter or pure-radiation), satisfying  Eqs. \eqref{Tref-cons} and \eqref{C-con}, is expected, together with \eqref{T-con},  to guarantee  
\eqref{papaefgenres} be Einstein. Scanning over canonical forms for $\text{T}^\pm$ 
amounts to exploring all Petrov classes. The Segre type of the reference tensors  determines precisely the Petrov type of the four-dimensional bulk metric and establishes a one-to-one map between the bulk Petrov type and the boundary data.\footnote{Some care must be taken when working with $\text{T}^\pm$ instead of $\text{S}^\pm$, because the eigenvalues are equal but not necessarily their eigenvectors. This means that one cannot determine the Petrov type unambiguously if considering the eigenvalue equation for $\text{T}^{\pm}$. The ambiguities occur between type II (III) and type D (N), since these types have the same degeneracy of eigenvalues. This was noticed e.g. in the Robinson--Trautman  metric studied in \cite{Gath:2015nxa}.}

Many examples illustrate the proposed pattern for producing exact four-dimensional algebraically special Einstein spaces from purely boundary considerations. These can be found in \cite{Gath:2015nxa, Petropoulos:2015fba}, and include families such as  Robinson--Trautman,  Pleba\'nski--Demia\'nski or Kundt spaces, for which the various Petrov classes are designed from the boundary data. 

In conclusion, Eqs.  \eqref{Tref-cons},  \eqref{C-con} and \eqref{T-con} appear as  \emph{a boundary translation of Einstein's equations, in the integrable sector of algebraically special geometries.} 
 
\subsection{A solution-generating transformation}\label{Ud}

Equations  \eqref{Tref-cons}, \eqref{C-con} and \eqref{T-con} are actually the \emph{holographic alternative of Geroch's three-dimensional sigma model} encoded in \eqref{gen-L-den}. 
The logic behind this statement is as follows. 

The starting point for the holographic determination of exact Einstein spaces is an ansatz for the boundary metric $\text{d}s^2_{\text{bry.}}$ (falling necessarily in the general class \eqref{PDbdymet}, but possibly expressed in a different way). The second step is the choice of a canonical form for $\text{T}^+$. At  this stage the conservation of $\text{T}^+$ \eqref{Tref-cons}, together with the equation for the Cotton \eqref{C-con}, allow to determine the actual boundary metric in conjunction with the adequate conserved reference tensors. From the latter, in a third step, one extracts the appropriate boundary energy--momentum tensor using \eqref{T-con}. This \emph{whole procedure is exclusively boundary-based}, and no other equation needs to be solved. The last step is the uplift to four dimensions, performed using \eqref{papaefgenres} with the boundary data at hand. 

In Geroch's programme, one would start with a reference metric $\hat{\text{h}}$ and determine  $(\kappa, \tau)$ by solving Eqs. \eqref{eq-tau} and \eqref{eq-kap} together with the constraints stemming out of Eqs. \eqref{eq-hab}. These three-dimensional data, namely $(\hat{\text{h}},\kappa, \tau)$ equivalent to $(\text{h},\omega,\lambda)$, would then be promoted to a four-dimensional Einstein space, following the pattern described in Sec. \ref{3to4}. 

This similarity between the two approaches can be pushed further, at the level of solution-generating methods. For Geroch, the procedure was recalled in Sec. \ref{integr}, and captured in \eqref{solgen-GL}
for Einstein spaces. In the holographic approach, we  observe that the conservation condition of the reference tensor, Eq. \eqref{Tref-cons}, whose key r\^ole in the integrability has been emphasized above, is invariant under the transformation
\begin{equation}
\label{hol-gen}
\text{T}^+\to z\, \text{T}^+, \quad \text{T}^-\to \bar z\, \text{T}^-, \quad z\in \mathbb{C}.
    \end{equation}
Any solution of the boundary integrability equations \eqref{Tref-cons},  \eqref{C-con} and \eqref{T-con}, can thus be mapped onto another one, potentially new. The relevance of this statement resides in the complex nature of the parameter $z$. Indeed, the transformation \eqref{hol-gen} induces a drastic modification of the boundary data: the energy--momentum and the Cotton tensors transform as
\begin{equation}
\label{hol-gen-cotem}
\text{T}\to \frac{1}{2} \left( z\, \text{T}^++\bar z\, \text{T}^-\right),\quad 
\text{C}\to \frac{3 k^3}{2i\kappa} \left( z\, \text{T}^+-\bar z\, \text{T}^-\right),
    \end{equation}
and the latter alters substantially the boundary metric.      
The uplifted four-dimensional  space \eqref{papaefgenres} is transformed accordingly into another Einstein geometry.

It is important to stress the local nature of the transformation \eqref{hol-gen} at the level of the reference tensors $\text{T}^\pm$. This transformation is also local for the actual energy--momentum tensor $\text{T}$, but \emph{non-local} for the boundary and for the resummed bulk  metrics. This latter statement is not surprising, as the transformation at hand acts ultimately on the bulk Weyl tensor, mixing its self-dual and anti-self-dual components $\text{Q}^{\pm}$, the boundary image of which being $\text{T}^\pm$ (more precisely
$\text{S}^\pm=-\text{P}^{-1}
\text{T}^\pm
\text{P}
$).

Before closing this chapter, we would like to elaborate on a specific canonical form for the boundary reference tensors $\text{T}^{\pm}$, namely the perfect-fluid form, which accounts for Petrov-D and Petrov-II bulk Einstein spaces (see Refs. \cite{Gath:2015nxa, Petropoulos:2015fba}):
\begin{equation}
\label{pfrefT}
\text{T}^\pm_{\text{pf}}=p_\pm(x)\left(3\left(\text{u}^\pm\right)^2 +\text{d}s^2_{\text{bry.}}\right).
\end{equation}
In this expression, $x$ stands for the  boundary coordinates $(t,\zeta,\bar \zeta)$ used in the generic form \eqref{PDbdymet}, and the two reference velocities are\footnote{These are the most general ones: adding an extra leg along the missing direction, and adjusting the overall scale for keeping the norm to $-1$ amounts to a combination of a diffeomorphism and a Weyl transformation.}
 \begin{equation}
 \label{upmPDtzzb}
 \text{u}^+= \text{u}+ \frac{\alpha^+}{P^2}\text{d}\zeta, \quad
 \text{u}^-= \text{u}+ \frac{\alpha^-}{P^2}\text{d}\bar\zeta
  \end{equation}
with functions $\alpha^\pm(x)$ adjusted in conjunction with the pressure fields $p_\pm(x)$ 
 for ensuring the conservation of $\text{T}^\pm$ \emph{i.e.}  Euler's equations.  In 3 spacetime dimensions these read:
\begin{equation}
\label{PMP-Euler0}
 \begin{cases}
2\text{u}_\pm (\ln p_\pm)+3\, \Theta_\pm=0 \\
 \text{u}_\pm (\ln p_\pm)\, \text{u}^\pm 
+\text{3}\ln p_\pm+3\, \text{a}_\pm=0
\end{cases}
\end{equation}
with 
$\text{u}_\pm (f)=u_\pm^\mu \partial_\mu f$.
Combining these equations, we obtain:
\begin{equation}\label{Euler0-int}
\text{A}^\pm+\text{d}\ln p_\pm^{\nicefrac{1}{3}}=0,
\end{equation}
where the Weyl connection $\text{A}^\pm$ is expressed in terms of the velocity $\text{u}^\pm$, the acceleration $ \text{a}^\pm$ and the expansion $\Theta_\pm$ as in \eqref{Wcon}.
Equation \eqref{Euler0-int} is integrable if the Weyl connection $\text{A}^\pm$ is closed (hence locally exact). 
If $\text{dA}^\pm\neq 0$ the ``fluid'' flowing on $\text{u}^\pm $ is not perfect. If $\text{A}^\pm$ vanishes, it is perfect and isobar. 

The reference pressures $p_\pm(x)$ can be expressed in terms of $\varepsilon(x)$ and $c(x)$ defined in Eqs. \eqref{Piprop} and  \eqref{cofx}, by computing $T^\pm_{\mu\nu}u^\mu u^\nu$, where $u^\mu$ are the components of the hydrodynamic congruence \eqref{ut}:
\begin{equation}
\label{ppm}
T^\pm_{\mu\nu}u^\mu u^\nu=2 p_\pm(x)=
 \varepsilon(x)\pm i \frac{\kappa}{3 k^3}
 c(x).
\end{equation}
The invariance of \eqref{Tref-cons} under \eqref{hol-gen},  is recast here as the invariance of \eqref{Euler0-int} under 
\begin{equation}
\label{hol-perfluid}
p_+(x)\to z p_+(x), \quad p_-(x)\to \bar z p_-(x), \quad z\in \mathbb{C},
    \end{equation}
or equivalently 
\begin{equation}
\label{hol-perfluid-epsc}
 \begin{pmatrix}     \varepsilon \\  \frac{\kappa}{3 k^3} c \end{pmatrix}  
\to \vert z\vert  
\begin{pmatrix} 
\cos \psi & -\sin \psi
\\
\sin \psi & \cos \psi 
 \end{pmatrix}  
 \begin{pmatrix}     \varepsilon \\  \frac{\kappa}{3 k^3} c \end{pmatrix} 
 \end{equation}
with $z=\vert z \vert (\cos \psi + i \sin \psi)$. This is a rescaling combined with a rotation in the plane of 
boundary data $(\varepsilon,  \frac{\kappa}{3 k^3} c)$. 

In the bulk, the transformation \eqref{hol-perfluid-epsc} affects directly the $\Psi_2$ component of the Weyl tensor. Indeed, the boundary congruence $\text{u}$ on which $\text{T}^\pm$ is projected in \eqref{ppm}, is lifted in the bulk onto a principal null direction of the Weyl (see discussion in Sec. \ref{der-exp}), and enters the definition of $\Psi_2$. At large $r$ the latter behaves as
 \begin{equation}
\label{genpsi2}
\Psi_2\approx-\frac{3}{2 k \kappa r^3}\left(
\varepsilon(x) +  \frac{i\kappa}{3k^3}c(x)\right).
 \end{equation}
Owing to the fact that the real part of $\Psi_2$ is generically associated with the bulk mass, whereas the imaginary part corresponds to the bulk nut (or rotation), \emph{the solution-generating transformation \eqref{hol-perfluid-epsc} amounts to a rotation in the $(m,n)$ plane}, besides an overall rescaling. As recalled in the introduction, this is precisely what Geroch's $SL(2,\mathbb{R})$ does in Ricci-flat spaces, and fails to achieve for Einstein geometries (see Sec. \ref{integr} and Ref. \cite{Leigh:2014dja}).

\subsection[A concrete example]{A concrete example\footnote{The interested reader can find all details for the holographic reconstruction of the Pleba\'nski--Demia\'nski family in Ref. \cite{Petropoulos:2015fba}.}}

The complete Petrov-D, two-Killing set of Einstein spaces is known as Pleba\'nski--Demia\'nski family\cite{Plebanski:1976gy} (see also \cite{PMP-GP}). It can be obtained from purely boundary considerations, following the method described in Secs. \ref{der-exp} and \ref{rescon}.

Although appropriate for expressing the resummed bulk metric \eqref{papaefgenres}, the form \eqref{PDbdymet}  for the boundary metric is not convenient for implementing the two-Killing symmetry requirement. We will therefore parameterize differently the boundary metric, adapting two coordinates $\tau$ and $\varphi$ to the two commuting Killing fields, and letting $\chi$ be the third one. Up to an arbitrary conformal factor, which plays no r\^ole in holographic issues, such a metric can be expressed in terms of two arbitrary functions $F(\chi)$ and $G(\chi)$ as follows:
\begin{equation}
\label{PDbry}
\text{d}s^2=
-\frac{F -\chi^4 G }{F+G}
\text{d}\varphi^2 +
\frac{G -\chi^4 F }{F+G}\text{d}\tau^2
+2\chi^2\, \text{d}\varphi\, \text{d}\tau+\frac{\text{d}\chi^2}{F G }.
 \end{equation}

As our aim is to build Petrov-D bulk metrics, 
the reference energy--momentum tensors are chosen of perfect-fluid form, which is Segre type D.\footnote{Similarly, we can proceed with a more general ansatz and recover two-Killing Einstein spaces
other than Petrov-D, such as Kundt spaces.} We need for this an ansatz for two complex-conjugate, normalized congruences with
exact Weyl connection. Thanks to the presence of the two Killing fields, it is easy to design such congruences by normalizing a linear combination of these fields:
\begin{equation}
\label{upmPD}
\text{u}_{\pm}
=
\frac{\partial_\tau\pm i \partial_\varphi}{\chi^2\mp i}
\leftrightarrow
\text{u}^{\pm}=\frac{\pm i}{F+G}\left(
G\left(\text{d}\tau+\chi^2 \text{d}\varphi\right)
\pm i 
F\left(\chi^2\text{d}\tau -\text{d}\varphi\right)
\right).
 \end{equation}
These are non-expanding and accelerating with exact
Weyl connections
\begin{equation}
\label{Apm}
\text{A}^{\pm}=
\text{d}
\ln(\chi^2\mp i).
 \end{equation}
The corresponding perfect-fluid reference tensors, of the form \eqref{pfrefT}, are thus conserved with a $\chi$-dependent pressure obtained by integrating Eq. \eqref{Euler0-int} with \eqref{Apm}:
\begin{equation}
\label{Ppm}
p_\pm(\chi) =-\frac{\kappa k}{3}\frac{m\mp i n}{(\chi^2\mp i)^3}.
\end{equation}

The parameters $m$ and $n$ are arbitrary and survive all the way up to the bulk metric, where they appear as the mass and the nut charge in appropriate normalizations. On the boundary, they emerge as first integrals of \eqref{Tref-cons}, resulting from the symmetry \eqref{hol-gen}, or equivalently \eqref{hol-perfluid}. Under the latter transformations with $z = \vert z\vert (\cos \psi +i \sin \psi)$, these parameters are mapped as follows: 
\begin{equation}
\label{hol-perfluid-mn}
 \begin{pmatrix}     m \\  n \end{pmatrix}  
\to \vert z\vert  
\begin{pmatrix} 
\cos \psi & \sin \psi
\\
-\sin \psi & \cos \psi 
 \end{pmatrix}  
 \begin{pmatrix}     m \\ n \end{pmatrix}. 
 \end{equation}
As advertised in Sec. \ref{Ud}, this amounts to a rotation in the $(m,n)$ plane, accompanied with an overall rescaling. This is a non-local transformation in the bulk Einstein space, which will be presented below.
Notice that for $\psi=\nicefrac{-\pi}{2}$ and $\vert z\vert =1$, the transformation at hand is a gravitational duality map, 
known to exchange the mass and the nut charge: $(m,n)\to (-n,m)$. This is settled more generally in Ricci-flat spaces, where the duality maps the Riemann tensor to its dual. As for the full continuous $U$-duality group, this $\mathbb{Z}_2$ subgroup acts non-locally on the four-dimensional metric. The holographic language seems to be the appropriate one for handling these duality issues in the presence of a cosmological constant. 

The Cotton tensor can be computed for the general boundary ansatz \eqref{PDbry}. The resummability condition \eqref{C-con} is then imposed using the reference tensors \eqref{pfrefT}, and it results in third-order differential equations for the functions $F(\chi)$ and $G(\chi)$. These equations turn out to be tractable, and we find:
\begin{eqnarray}
\label{F}
&&F(\chi)=\frac{\hat R(\chi)}{\chi^4+1}=\frac{1}{\chi^4+1}\left[(k^2-\ell)\chi^4-2n\chi^3+\epsilon \chi^2-2m\chi +\ell\right],\\
\label{G}
&&G(\chi)= \frac{\hat Q(\chi)}{\chi^4+1}=\frac{1}{\chi^4+1}\left[\ell\chi^4+2n\chi^3-\epsilon \chi^2+2m\chi +k^2-\ell\right],
\end{eqnarray}
obeying
\begin{equation}
\label{FG}
F(\chi)+G(\chi)
= k^2
 \end{equation}
 with $\ell, \epsilon$ extra arbitrary integration constants.

As stressed in the general presentation of Sec. \ref{rescon}, the hydrodynamic congruence $\text{u}$ is part of the resummation ansatz. Once the boundary metric is set in the form \eqref{PDbdymet}, the velocity field used in the resummation formula \eqref{papaefgenres} should be taken to be \eqref{ut}, which is shear-free. It is possible to turn \eqref{PD} onto \eqref{PDbdymet} by trading $(\tau,\chi,\varphi)$ for $(t, \zeta, \bar\zeta)$ as\footnote{Equation  \eqref{FG} is used, even though $F$ and $G$ appear separately in several expressions.}
\begin{eqnarray}
\label{chi}
t+\frac{\zeta+\bar\zeta}{k\sqrt{2}}&=&\int\text{d}\chi\left(
\frac{k^2\chi^2}{FG(\chi^4+1)}-\frac{1}{F\chi^2}
\right),
\\
\label{zeta}
\zeta&=&-\frac{k}{\sqrt{2}}
\left(\tau+i\varphi-k^2\int\text{d}\chi\frac{\chi^2-i}{FG\left(\chi^4+1\right)}\right).
\end{eqnarray}
In the new frame, the two Killing vector fields read:
 \begin{equation}
\partial_{\tau}=\partial_t-\frac{k}{\sqrt{2}}
\left(\partial_\zeta+
\partial_{\bar \zeta}
\right),
\quad
\partial_{\varphi}
=-
i\frac{k}{\sqrt{2}}
\left(\partial_\zeta-
\partial_{\bar \zeta}
\right),
\label{Ktauphi}
\end{equation}
and $\partial_t$ \emph{is not} a Killing. Similarly, the complex-conjugate congruences \eqref{upmPD} are recast in the form \eqref{upmPDtzzb}  with appropriate $\alpha^{\pm} =\alpha^{\pm}(\chi)
$. It is worth reminding that these congruences are actually the most general ones, and this is why the choice at hand allows to reconstruct at the end the most general type-D, two-Killing Einstein space.

The boundary metric has now the form \eqref{PDbdymet}, and 
all functions $\Omega$, $P$ and $B$ depend only on $\chi$ \emph{i.e.} on the specific combination dictated by the isometries, $t+\frac{\zeta+\bar\zeta}{k\sqrt{2}}$. Expressed in the frame $(\text{d}\tau, \text{d}\chi, \text{d}\varphi)$, the 
hydrodynamic congruence $\text{u}$ given in \eqref{ut} reads:
\begin{equation}
\text{u}=\text{d}\varphi-
\chi^2\text{d}\tau +\frac{\text{d}\chi}{F}.\label{ uuPD}
 \end{equation}
Given the hydrodynamic congruence, it is possible to determine the longitudinal projections of the  energy--momentum and Cotton tensors, as displayed in Eqs. \eqref{Piprop} and \eqref{cofx}. We find:
 \begin{eqnarray}
 \label{epsPD}
\varepsilon 
(\chi)&=&-\frac{2k\kappa}{3}\frac{m\chi^2(\chi^4-3)+n(3\chi^4-1)}{(1+\chi^4)^3},\\
 \label{cPD}
c(\chi)&=&2k^4\frac{n\chi^2(\chi^4-3)-m(3\chi^4-1)}{(1+\chi^4)^3}.
 \end{eqnarray}
 
 At this stage we can perform the resummation of the derivative expansion associated with the above boundary data, using Eq. \eqref{papaefgenres}. Various  
intermediate ingredients are computed with Eqs. \eqref{sigma},  \eqref{curlR} and  \eqref{rho2}. 
It is wise to abandon the analogue of Eddington--Finkelstein coordinates $(\tau, \chi, \varphi, r)$ in use, and bring the metric at hand in a more familiar form. This is achieved with the following coordinate transformation:
\begin{equation}
\label{trans}
\begin{cases}
\tau&=\hat \tau+\int \frac{\hat{q}^2\, \text{d}\hat q}{\hat R(\hat q)}-\int 
\frac{\hat{r}^2\, \text{d}\hat r}{\hat R(\hat r)}
\\
\chi&=\hat q
\\
\varphi&=\hat \varphi-\int \frac{\text{d}\hat q}{\hat R(\hat q)}+\int 
\frac{\text{d}\hat r}{\hat R(\hat r)}
\\
r&=\frac{1}{\hat r-\hat q}+\frac{\hat{q}^3}{\hat{q}^4+1}
,
\end{cases}
\end{equation}
where the function $\hat R$ is displayed in \eqref{F}. Despite the complicated expression of the original metric, 
thanks to the polynomial structure of $\hat R$ and to Eqs. \eqref{F}, \eqref{G} and \eqref{FG}, the metric  \eqref{papaefgenres} becomes unexpectedly simple  in
the new coordinates $(\hat\tau, \hat q, \hat\varphi, \hat r)$:
\begin{eqnarray}
\text{d}s^2_{\text{bry.}}&=&\frac{1}{(\hat{q}-\hat{r})^2}\bigg(
-\frac{\hat R(\hat{r}) \left(\text{d}\hat{\varphi}-\hat{q}^2\text{d}\hat{\tau}\right)^2}{1+(\hat{r}\hat{q})^2}
+\frac{\hat Q(\hat{q}) \left(\text{d}\hat{\tau}+\hat{r}^2\text{d}\hat{\varphi}\right)^2}{1+(\hat{r}\hat{q})^2}
\nonumber
\\
\label{PD}
&&+
\frac{1+(\hat{r}\hat{q})^2}{\hat Q(\hat{q})} \text{d}\hat{q}^2
+
\frac{1+(\hat{r}\hat{q})^2}{\hat R(\hat{r})} \text{d}\hat{r}^2\bigg).
 \end{eqnarray}
This is the Pleba\'nski--Demia\'nski  geometry  with zero electric and magnetic charges  \cite{Plebanski:1976gy}. It is Einstein and Petrov-D, and its unique Weyl invariant is 
\begin{equation}
\Psi_2=(m-in)\left(
\frac{\hat{q}-\hat{r}}{i-\hat{r}\hat{q}}
\right)^3.
 \end{equation}
At large $r$, the latter behaves as advertised in \eqref{genpsi2} with $\varepsilon$ and $c$ displayed in \eqref{epsPD} and \eqref{cPD}.
 
\section*{Conclusion}
\addcontentsline{toc}{section}{Conclusions}

The Geroch group emerges as a symmetry of the three-dimensional sigma model obtained when reducing four-dimensional vacuum Einstein's equations along a Killing congruence. This group is at the heart of remarkable integrability properties, and acting on a solution, it provides another solution. The whole scheme is extendable to higher dimensions, where it leads to the $U$-duality groups, very popular in supergravity and superstring compactifications. 

When considering Einstein spaces instead of Ricci-flat geometries, the Geroch group still exists, but not all of its generators produce solutions. It is therefore unclear whether it accounts for the integrability properties, which are still present though in some corners of the phase space. An alternative perspective for understanding these properties and setting up solution-generating techniques is made available within holography.

In holography, expansions such as the Fefferman--Graham, or the more physical hydrodynamic derivative expansion are resummable under appropriate conditions on the boundary data, which are the boundary metric and the boundary energy--momentum tensor. Displaying these conditions and exploring 
their invariance has been the core of this presentation.

The analysis we performed is rewarding, as integrability conditions do exist and are invariant under an operation that produces non-trivial transformations on the boundary data, as well as on the bulk geometry. The whole procedure articulates around the complex-conjugate reference tensors $\text{T}^\pm$. These are
\begin{itemize}
\item of canonical Segre form, because they are the holographic image of the bulk Weyl tensor (up to a similarity transformation), itself algebraically special due to the assumption of shear-free hydrodynamic congruence $\text{u}$; 
\item conserved, $\nabla\cdot \text{T}^\pm=0$, and thus freely transforming under 
$\text{T}^+\to z\, \text{T}^+$, $ \text{T}^-\to \bar z\, \text{T}^-$ with constant  $z\in \mathbb{C}
$.
\end{itemize}
The integrability conditions  \eqref{Tref-cons}, \eqref{C-con} and \eqref{T-con} are  the 
boundary manifestation of Einstein bulk equations in some integrable sector. They are the analogue of Geroch's three-dimensional sigma model \eqref{gen-L-den}, and their invariance quoted previously replaces the $N\subset SL(2,\mathbb{R})$ invariance group of \eqref{gen-L-den}. This invariance appears as a boundary-controlled holographic $U$-duality, ultimately mixing the self-dual and anti-self-dual components $\text{Q}^\pm$ of the bulk Weyl tensor.  The fact that this mixing is achieved with the help of the Cotton and 
the energy--momentum tensor is not surprising. What is remarkable, is that integrability is preserved, producing a genuine solution-generating pattern, non-locally relating Einstein spaces. The Weyl gravitational duality is a $\mathbb{Z}_2$ subgroup of the holographic $U$-duality symmetry.

Understanding the deeper relationship that may exist between the Geroch and the holographic symmetries, setting a bridge between Fefferman--Graham and hydrodynamic derivative expansions, and extending the whole scheme to higher dimensions are related and challenging questions that deserve further investigation.

\section*{Acknowledgements}

This work is inspired by a talk delivered during the conference \textsl{Planck 2015} held in May 2015 at the University of Ioannina, Greece. 
It relies on works published or to appear, performed in collaboration with M. Caldarelli, J.~Gath, R. Leigh, A. Mukhopadhyay, V. Pozzoli and P. Tripathy. The work of Anastasios Petkou is partially supported by the research
grant ARISTEIA II, 3337,  \textsl{Aspects of three-dimensional CFTs}, by the Greek General
Secretariat of Research and Technology, and also by the CreteHEPCosmo-228644 grant.
The research of P.M. Petropoulos and K.~Siampos is supported by the Franco--Swiss bilateral Hubert Curien program  \textsl{Germaine de Stael} 2015 (project no 32753SG). We thank each others home institutions for hospitality and financial support.


\begin{thebibliography}{99.}%

\bibitem{Geroch}
R. Geroch, \textit{A method for generating solutions of Einstein's equations,} 
\href{http://scitation.aip.org/content/aip/journal/jmp/12/6/10.1063/1.1665681}{J. Math Phys. \textbf{12} (1971)~918.}

\bibitem{Geroch:1972yt}
  R.~Geroch,
  \textit{A method for generating new solutions of Einstein's equation, II,}
  \href{http://scitation.aip.org/content/aip/journal/jmp/13/3/10.1063/1.1665990}{J.\ Math.\ Phys.\  {\bf 13} (1972) 394.}

\bibitem{Ehlers} J. Ehlers, \textsl{Les th\'eories relativistes de la gravitation}, CNRS, Paris, 1959, 275-284.

  
\bibitem{Alekseev:1980ew} 
  G.A.~Alekseev,
  \textit{$N$ soliton solutions of Einstein--Maxwell equations},
\href{http://www.jetpletters.ac.ru/ps/1426/article_21693.shtml}{JETP Lett. {\bf 32} No. 4 
 (1980) 20.} 
  

\bibitem{Mazur:1983}
P.O. Mazur, \textit{A relationship between the electrovacuum Ernst equations and nonlinear sigma model}, 
\href{http://www.actaphys.uj.edu.pl/_old/vol14/abs/v14p0219.htm}{Acta Phys. Polon. \textbf{B14} (1983) 219.}
  
\bibitem{Alekseev:2010mx} 
  G.A.~Alekseev,
  \textit{Thirty years of studies of integrable reductions of Einstein's field equations},
  \href{http://arxiv.org/abs/1011.3846}{arXiv:1011.3846 [gr-qc].}

\bibitem{Ernst:1967wx} 
  F.J.~Ernst,
  \textit{New formulation of the axially symmetric gravitational field problem,}
  \href{http://journals.aps.org/pr/abstract/10.1103/PhysRev.167.1175}{Phys.\ Rev.\  {\bf 167},  (1968) 1175.
  }
  
 \bibitem{Ernst:1968} 
  F.J. Ernst, \textit{New formulation of the axially symmetric gravitational field problem, II,} 
  \href{http://journals.aps.org/pr/abstract/10.1103/PhysRev.168.1415}{Phys. Rev. \textbf{168} (1968) 1415.}

\bibitem{Breitenlohner:1986um}
  P.~Breitenlohner and D.~Maison,
  \textit{On the Geroch group,}
  \href{https://eudml.org/doc/76358}{Annales IHP \  {\bf A46} (1987) 215.}
  
  \bibitem{belinskii}
 V. Belinskii and V. Zakharov, 
 \textit{Integration of the Einstein equations by means of the inverse scattering problem technique and construction of exact soliton solutions,}\\
  \href{http://www.jetp.ac.ru/cgi-bin/e/index/e/48/6/p985?a=list}{Sov. Phys. JETP \textbf{48}  (1978) 6.}
  
\bibitem{Maison:1978es}
  D.~Maison,
  \textit{Are the stationary, axially symmetric Einstein equations completely integrable?,}
  \href{http://journals.aps.org/prl/abstract/10.1103/PhysRevLett.41.521}{Phys.\ Rev.\ Lett.\  {\bf 41} (1978) 521.}
  
   \bibitem{maison2}
D. Maison,
  \textit{On the complete integrability of the stationary, axially symmetric Einstein equations,}
\href{http://scitation.aip.org/content/aip/journal/jmp/20/5/10.1063/1.524134}{J. Math. Phys. \textbf{20} (1979) 871.}

   
  \bibitem{Mazur:1982}
P.O. Mazur,  \textsl{Properties and integrability of the Ernst equations}, Ph. D. Thesis, Jagellonian University, Krakow, Poland, unpublished (in Polish).

\bibitem{Bernard:2001pp}
  D.~Bernard and N.~Regnault,
  \textit{New Lax pair for 2D dimensionally reduced gravity},
  \href{http://iopscience.iop.org/article/10.1088/0305-4470/34/11/325/pdf;jsessionid=11EB55E7BD2F22C453E22C51B0E25C8D.c2.iopscience.cld.iop.org}{J.\ Phys. {\bf A34} (2001) 2343.}

\bibitem{Bardoux:2013swa}
  Y.~Bardoux, M.M.~Caldarelli and C.~Charmousis,
  \textit{Integrability in conformally coupled gravity: Taub--NUT spacetimes and rotating black holes},
  JHEP {\bf 1405} (2014) 039,
  \href{http://arxiv.org/abs/1311.1192}{arXiv:1311.1192 [hep-th].}

\bibitem{Alekseev:2004zz} 
  G.A.~Alekseev,
  \textit{Integrability of generalized (matrix) Ernst equations in string theory},
  Theor.\ Math.\ Phys.\  {\bf 144} (2005) 1065 
  [Teor.\ Mat.\ Fiz.\  {\bf 144} (2005) 214],
  \href{http://arxiv.org/abs/hep-th/0410246}{arXiv:hep-th/0410246.}
  
\bibitem{Alekseev:2008gh} 
  G.A.~Alekseev,
  \textit{Integrability of the symmetry reduced bosonic dynamics and soliton generating transformations in the low-energy heterotic string effective theory},
  Phys.\ Rev.\ {\bf D80} (2009) 041901,
  \href{http://arxiv.org/abs/0811.1358}{arXiv:0811.1358 [hep-th].}
  
\bibitem{Figueras:2009mc} 
  P.~Figueras, E.~Jamsin, J.V.~Rocha and A.~Virmani,
  \textit{Integrability of five dimensional minimal supergravity and charged rotating black holes},
  Class.\ Quant.\ Grav.\  {\bf 27} (2010) 135011,
  \href{http://arxiv.org/abs/0912.3199}{arXiv:0912.3199 [hep-th].}
  
\bibitem{Mishima:2005id} 
  T.~Mishima and H.~Iguchi,
  \textit{New axisymmetric stationary solutions of five-dimensional vacuum Einstein equations with asymptotic flatness},
  Phys.\ Rev.\  {\bf D73} (2006) 044030,
  \href{http://arxiv.org/abs/hep-th/0504018}{hep-th/0504018.}


\bibitem{Iguchi:2007is} 
  H.~Iguchi and T.~Mishima,
  \textit{Black di-ring and infinite nonuniqueness},
  Phys.\ Rev.\ {\bf D75} (2007) 064018,
  [Erratum-ibid.\  {\bf D78} (2008) 069903],
  \href{http://arxiv.org/abs/hep-th/0701043}{hep-th/0701043.}



\bibitem{Charmousis:2006fx} 
  C.~Charmousis, D.~Langlois, D.A.~Steer and R.~Zegers,
  \textit{Rotating spacetimes with a cosmological constant,}
  JHEP {\bf 0702} (2007) 064,
  \href{http://arxiv.org/abs/gr-qc/0610091}{gr-qc/0610091.}

\bibitem{Caldarelli:2008pz} 
  M.M.~Caldarelli, R.~Emparan and M.J.~Rodriguez,
  \textit{Black rings in (anti-)de Sitter space,}
  JHEP {\bf 0811} (2008) 011,
  \href{http://arxiv.org/abs/0806.1954}{arXiv:0806.1954 [hep-th].}
  

\bibitem{Astorino:2012zm}
  M.~Astorino,
  \textit{Charging axisymmetric space-times with cosmological constant,}
  JHEP {\bf 1206} (2012) 086,
  \href{http://arxiv.org/abs/1205.6998}{arXiv:1205.6998 [gr-qc].}
  
\bibitem{Leigh:2014dja}
  R.G.~Leigh, A.C.~Petkou, P.M.~Petropoulos and P.K.~Tripathy,
  \textit{The Geroch group in Einstein spaces,}
  Class.\ Quant.\ Grav.\  {\bf 31} (2014)  225006,
  \href{http://arxiv.org/abs/1403.6511}{arXiv:1403.6511 [hep-th].}

\bibitem{PMP-FG1}
C. Fefferman and C.R. Graham, \textit{Conformal invariants}, in \textsl{Elie Cartan et les math\'ematiques d'aujourd'hui,} Ast\'erisque, 1985, num\'ero hors s\'erie Soc. Math. France, Paris, 95.

\bibitem{PMP-FG2}
C. Fefferman and C.R. Graham, \textit{The ambient metric}, \href{http://arxiv.org/abs/0710.0919}{arXiv:0710.0919 [math.DG].}

\bibitem{Caldarelli:2012cm}
M.M. Caldarelli, R.G. Leigh, A.C. Petkou, P.M. Petropoulos, V. Pozzoli and K. Siampos, 
\textit{Vorticity in holographic fluids},   Proc. of Science \textbf{Corfu11} (2012) 076,
\href{http://arxiv.org/abs/1206.4351}{arXiv:1206.4351 [hep-th].}

\bibitem{Mukhopadhyay:2013gja}
  A.~Mukhopadhyay, A.C.~Petkou, P.M.~Petropoulos, V.~Pozzoli and K.~Siampos,
  \textit{Holographic perfect fluidity, Cotton energy--momentum duality and transport properties},
  JHEP \textbf{04} (2014) 136, \href{http://arxiv.org/abs/1309.2310}{arXiv:1309.2310 [hep-th].}
    
  \bibitem{Petropoulos:2014yaa}
P.M.~Petropoulos,
\textit{Gravitational duality, topologically massive gravity and holographic fluids},
Lect.\ Notes Phys.\  {\bf 892} (2015) 331,
\href{http://arxiv.org/abs/1406.2328}{arXiv:1406.2328 [hep-th]}. 

\bibitem{Gath:2015nxa}
  J.~Gath, A.~Mukhopadhyay, A.C.~Petkou, P.M.~Petropoulos and K.~Siampos,
  \textit{Petrov classification and holographic reconstruction of spacetime,}
  JHEP {\bf 1509} (2015) 005,
  \href{http://arxiv.org/abs/1506.04813}{arXiv:1506.04813 [hep-th].}

\bibitem{Petropoulos:2015fba}
  P.M.~Petropoulos and K.~Siampos,
   \textit{Integrability, Einstein spaces and holographic fluids},
    \href{http://arxiv.org/abs/1510.06456}{arXiv:1510.06456 [hep-th].}
  
\bibitem{Berkeley:2015mmc}
  J.A.~Fitzhardinge-Berkeley,
 \textit{Solution-generating transformations in duality-invariant theories and the fluid/gravity correspondence}, \href{http://arxiv.org/abs/1511.00995}{arXiv:1511.00995 [hep-th]}

\bibitem{Bhattacharyya:2008jc}
  S.~Bhattacharyya, R.~Loganayagam, I. Mandal, S.~Minwalla and A. Sharma,
  \textit{Conformal nonlinear fluid dynamics from gravity in arbitrary dimensions},
  JHEP {\bf 0812} (2008) 116,
  \href{http://arxiv.org/abs/0809.4272}{arXiv:0809.4272 [hep-th].}

\bibitem{Bhattacharyya:2008ji}
  S.~Bhattacharyya, R.~Loganayagam, S.~Minwalla, S.~Nampuri, S.P.~Trivedi and S.R.~Wadia,
  \textit{Forced fluid dynamics from gravity},
  JHEP {\bf 0902} (2009) 018,
  \href{http://arxiv.org/abs/0806.0006}{arXiv:0806.0006 [hep-th].}
 
\bibitem{Haack:2008cp}
  M.~Haack and A.~Yarom,
  \textit{Nonlinear viscous hydrodynamics in various dimensions using AdS/CFT,}
  JHEP {\bf 0810} (2008) 063,
  \href{http://arxiv.org/abs/0806.4602}{arXiv:0806.4602 [hep-th].}
 
 \bibitem{PMP-GP}
J. Griffiths  and J. Podolsk\' y, \textsl{Exact spacetimes  in Einstein's general relativity},  
Cambridge Monographs on Mathematical Physics,  CUP 2009.

\bibitem{Coll}
  B.~Coll, J.~Llosa and D.~Soler,
  {\it Three-dimensional metrics as deformations of a constant curvature metric},
  Gen.\ Rel.\ Grav.\  {\bf 34} (2002) 269,
  \href{http://arxiv.org/abs/gr-qc/0104070}{gr-qc/0104070.}

\bibitem{PMP-LeBrun82}
C.R. Lebrun, \textit{$\mathcal{H}$-space with a cosmological constant}, 
\href{http://rspa.royalsocietypublishing.org/content/380/1778/171}{Proc. R. Soc. Lond. \textbf{A380} (1982) 171.}


\bibitem{Stephani:624239}
H. Stephani, D. Kramer, M. MacCallum, C. Hoenselaers and E. Herlt, \textit{Exact solutions to Einstein's field equations},  Cambridge Monographs on Mathematical Physics,  CUP 2003.
 
\bibitem{Mansi:2008br} 
  D.S.~Mansi, A.C.~Petkou and G.~Tagliabue,
  \textit{Gravity in the $3+1$-split formalism I: holography as an initial value problem},
  Class.\ Quant.\ Grav.\  {\bf 26} (2009) 045008,
  \href{http://arxiv.org/abs/0808.1212}{arXiv:0808.1212 [hep-th].}
  
\bibitem{Mansi:2008bs} 
  D.S.~Mansi, A.C.~Petkou and G.~Tagliabue,
  \textit{Gravity in the $3+1$-split formalism II: self-duality and the emergence of the gravitational Chern--Simons in the boundary},
  Class.\ Quant.\ Grav.\  {\bf 26} (2009) 045009, 
  \href{http://arxiv.org/abs/0808.1213}{arXiv:0808.1213 [hep-th].}
  
\bibitem{Chow:2009km}
  D.D.K.~Chow, C.N.~Pope and E.~Sezgin,
  \textit{Classification of solutions in topologically massive gravity},
  Class.\ Quant.\ Grav.\  {\bf 27} (2010) 105001,
  \href{http://arxiv.org/abs/0906.3559}{arXiv:0906.3559 [hep-th].}
  
  
\bibitem{Plebanski:1976gy}
  J.F.~Pleba\'nski and M.~Demia\'nski,
  \textit{Rotating, charged, and uniformly accelerating mass in general relativity},
  \href{http://www.sciencedirect.com/science/article/pii/0003491676902402}{Annals Phys.\  {\bf 98} (1976) 98.}



\end{thebibliography}

\end{document}